\documentclass[aps,prl,twocolumn,superscriptaddress]{revtex4-1}

\usepackage{amssymb,amsfonts,amsmath}
\usepackage{euscript}
\usepackage{graphicx}
\usepackage{xcolor}
\usepackage{natbib}

\begin{document}

\title{Collective Dynamics of Belief Evolution under Cognitive Coherence and Social Conformity}
\author{Nathaniel Rodriguez} 
\author{Johan Bollen}
\author{Yong-Yeol Ahn}
%\footnote{Corresponding author: yyahn@indiana.edu}
\email[Corresponding author: ]{yyahn@indiana.edu}
\affiliation{The Center for Complex Networks and Systems Research, School of Informatics and Computing, Indiana University, Bloomington, Indiana, United States}

\begin{abstract}
Human history has been marked by social instability and
conflict, often driven by the irreconcilability of opposing sets of beliefs,
ideologies, and religious dogmas. 
The dynamics of belief systems has been studied mainly from two distinct perspectives, namely how cognitive biases lead to individual belief rigidity and how social influence leads to social conformity.
Here we propose a unifying framework that connects cognitive and
social forces together in order to study the dynamics of societal belief evolution.
Each individual is endowed with a network of interacting
beliefs that evolves through interaction with other individuals in a social network.
The adoption of beliefs is affected by both internal coherence and social
conformity. Our framework may offer explanations for how social transitions can arise in
otherwise homogeneous populations, how small numbers of zealots with highly
coherent beliefs can overturn societal consensus, and how belief rigidity 
protects fringe groups and cults against invasion from
mainstream beliefs, allowing them to persist and even thrive in larger
societies. Our results suggest that strong consensus may be
insufficient to guarantee social stability, that the cognitive coherence of
belief-systems is vital in determining their ability to spread, and that
coherent belief-systems may pose a serious problem for resolving social
polarization, due to their ability to prevent consensus even under high levels
of social exposure. We argue that the inclusion of cognitive factors
into a social model could provide a more complete picture of
collective human dynamics.
\end{abstract}

\maketitle

\section*{Introduction}

Ideological conflict has been a major challenge for human
societies~\cite{sowell-ideology_conflict-2002}.  For instance, when post
World-War I Germany was marked by economic depression and social trauma,
ideological fringe groups like the National-Socialist Party and the Communist
Party of Germany made their way into mainstream politics to eventually dominate
the political landscape ~\cite{hoffer-truebeliever-1966,
rohkramer-germany-2007, mcdonough-hitler-2012}. This process of ideological
upheaval eventually led to World War II, one of the deadliest conflicts in
human history. Similarly, 14th and 15th century Europe was torn by sharp
religious transitions and accompanying widespread conflicts, such as the Thirty
Years' War~\cite{britannica-reformation-2014}.  The abundance of such
ideological transitions in history raises questions: are they driven
by common psycho-social mechanisms? Do specific peculiarities of human
psychology play a role in ideological dynamics?

\begin{figure}[h!]
\centering
	\includegraphics[width=8.7cm]{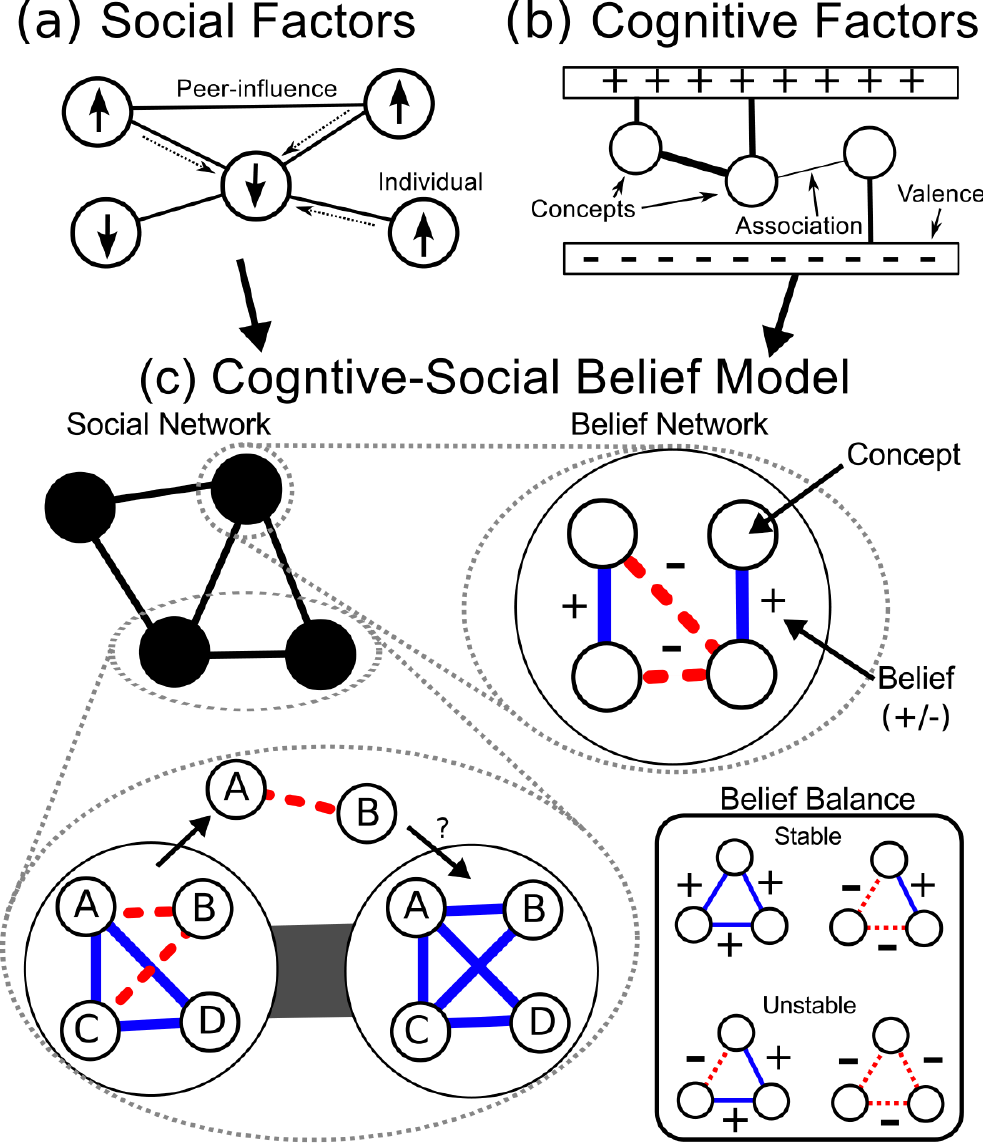}

	\caption{\label{Fig 1}\textbf{Cognitive-Social Belief Model} (a) Social models, such as the voter or
Sznajd models, focus on the assimilation process through social pressure.
Beliefs are usually simplified as independent states.  (b) Cognitive models,
such as the SKS model, focus on the interaction and coherence of beliefs of a
single individual and how individuals make decisions and change their minds.
The effect of social networks is often unaddressed.  (c) Our model incorporates
both forces, recognizing not only social pressures but also the connected
nature of human beliefs.  The social network acts as a conduit for belief
transmission between individuals.  We model a belief as a signed relationship
between two concepts. We express the internal coherence of a network of such
beliefs in terms of social balance theory where relationship triads can be
either stable or unstable.  The belief networks evolve over time as individuals
decide whether to accept new beliefs transmitted by their peers.} 

\end{figure}

Although these questions have been tackled by numerous studies, most existing
models of belief system dynamics focus on either social or cognitive factors
rather than integrating both aspects.  Social models focus on how social
interactions transmit and shape beliefs.  For instance, Axelrod's cultural
dissemination model considers social influence and homophily as key drivers of
cultural polarization (see Fig.~\ref{Fig 1}(a))~\cite{axelrod-culture-2009, hammond-ethnocentrism-2006}.  
In this model each agent is represented as a vector of independent traits that can be modified
through social influence. The study of spin systems in
physics have inspired a number of opinion models, such as the voter model~\cite{castellano-socdyn-2009,Masuda-voter-2010,sood-hetnetvoter-2007},
Sznajd model~\cite{sznajd-opinionmodel-2000,clifford-votermodel-1973,Javarone-conformity-2014}, and Ising-like models~\cite{Galam-opinion-2000}.
Other approaches have drawn upon reaction-diffusion systems~\cite{Galam-review-2008}, or may use continuous opinions~\cite{Guillaume-opinion-2000} or bounded-confidence~\cite{Hegselmann-opinion-2002}.

Cognitive modeling approaches mainly focus on information processing and
decision making (see
Fig.~\ref{Fig 1}(b))~\cite{Allahverdyan-ConfirmationBias-2014}.
Psychological research has revealed that individuals strive for internal
consistency, which leads to cognitive mechanisms such as \emph{confirmation
bias}~\cite{nickerson-confirmationbias-1998} and \emph{cognitive
dissonance}~\cite{festinger-dissonance-1962}. Our understanding of these cognitive
forces and biases in belief formation allow us to move beyond the underlying assumptions---a
set of independent beliefs~\cite{goldstone-collectiveBehavior-2005,
mason-socialpsycreview-2007}---of spin models or opinion vector-based
models~\cite{mcgarty-turnerexp-1992, pechey-sharedbeliefs-2012}. 

Recent attempts to combine these forces have used mechanisms like social- and
anti-conformity~\cite{Javarone-conformity-2014}, foundational beliefs~\cite{redner-truth-2011}, 
confirmation bias~\cite{nishi-socialconfbias-2013}, attitudes~\cite{Wang-opinion-2014},  and social network
formation~\cite{javarone-socialstructure-2013, Sekara-socialstructure-2015,
shi-coevolvingvotermodel-2013}. However, modeling belief-systems that
consist of interacting beliefs, and the study of such systems under social
influence, has not been fully investigated. Here we introduce a novel framework that can
incorporate both social and cognitive factors in a coherent way (see
Fig.~\ref{Fig 1}(c)). We show that the integration of social and
cognitive factors produces elementary features of collective social
phenomena---societal transitions, upheavals, and existence of fringe groups. 

Our framework represents a society as a network of individuals---a social
network---where each individual possesses a network of concepts and beliefs.
Social influence takes place via the social ties~\cite{castellano-socdyn-2009,
lord-attitude_polarize-1979, taber-politicalbel-2006, evans-beliefsys-xxxx} and
each individual carries a belief network of interconnected relationships that
represents the individual's belief system~\cite{festinger-dissonance-1962, pechey-sharedbeliefs-2012,greenwald-iat-2002, schilling-cognet-2005, quine-beliefweb-1970, thagard-coherence-2000,
pechey-beliefcoherence-2012}. We evaluate the internal coherence of each
individual's belief network by applying the balance
theory~\cite{heider-cog_organization-1946, norman-balancetheory_group-2003}.
The coherence shifts an individual's willingness to integrate new beliefs, thus
simulating cognitive traits such as confirmation bias and cognitive
dissonance~\cite{nickerson-confirmationbias-1998, mcgarty-turnerexp-1992,lord-attitude_polarize-1979, taber-politicalbel-2006,evans-beliefsys-xxxx, Allahverdyan-ConfirmationBias-2014}. At the same time, an individual's belief network
is affected by social influence; repeated exposure to new ideas through social
ties~\cite{asch-groupress-1951, asch-conformity-1956, centola-onlinenet-2010,
matz-cogdis_in_groups-2005} increases their likelihood of adopting even
incongruent ideas. Individuals thus experience both cognitive and social
forces: (i) they prefer to have coherent belief networks and prefer beliefs
that will increase their internal coherence; but (ii) may accept conflicting
beliefs under the influence of strong social pressure.

\section*{Methods}

Our approach considers a network of concepts and beliefs where the nodes
represent concepts and signed edges between them represents binary associative
beliefs that capture the relation between two concepts (cf. Social Knowledge
Structure (SKS) model~\cite{greenwald-iat-2002}).  This formulation allows us
to define internal coherence through the principle of triad stability in
social balance theory~\cite{heider-original-1946, antal-social_balance-2006,
doreian-social_balance-2002, cartwright-balance_theory-1957}.   

For instance, consider the beliefs of Alice, who is a devoted spectator of \emph{soccer}. The \emph{Eagles} are her favorite soccer team, but it has been charged with \emph{match fixing}. In Alice's belief network there is a positive link between the \emph{Eagles} and \emph{soccer}, and between the \emph{Eagles} and \emph{match fixing}, while there is a negative link between \emph{soccer} and \emph{match fixing}. Such
pressured configurations are considered unstable (incoherent), and are
analogous to frustrated states in spin-systems or unstable social triads. To
resolve the frustration Alice may dissociate the \emph{Eagles} from the allegations of match fixing, drop the \emph{Eagles}' association with the sport, or change the relationship between \emph{soccer} and \emph{match fixing}. It has been shown that people tend to quickly resolve such
inconsistency when provided the opportunity to choose dissonance reduction
strategies~\cite{elliot-cogdis-1994}.  Yet, in the presence of social pressure
or more complicated concept associations, a concept may remain pressured. Each
triad in a belief network can be either stable or unstable, as shown in
Fig.~\ref{Fig 1}(c). The incoherence of an entire belief system (of individual $n$) can thus
be captured by an internal energy function~\cite{marvel-social_balance_energy-2009}
on the belief network $\EuScript{M}$:

\begin{eqnarray}\label{eq:indive}
E_n^{(i)} = -\frac{1}{\binom{M}{3}} \sum_{j,k,l} a_{jk} a_{kl} a_{jl}, 
\end{eqnarray}
where $M$ is the number of nodes in the belief network and $a_{jk}$ is the association connecting nodes $j$ and $k$, which can be $+1$ (positive association) or $-1$ (negative
association).  The sum is taken over all triads in the belief network and normalized by the total number of triads. For simplicity,
in our simulations we choose this network to be complete, meaning that all 
concepts have a positive or negative association with every other concept.

A single association's contribution to the energy depends on the state of
adjacent associations, providing interdependence and rigidity to the belief
system.  Beliefs do not necessarily reflect reality.  They may be fabricated or
completely false.  It is, however, the \emph{interaction} between
beliefs that gives them their strength, reflective of psychological factors
like confirmation bias and cognitive dissonance.

The evolution of belief systems is also driven by social interactions, through
which people communicate their beliefs to others.  We represent this society
as a social network, $\EuScript{N}$, where $N=|\EuScript{N}|$, and 
whose nodes are individuals and edges represent social
relationships through which ideas are communicated.  We define a second,
``social'' energy term, inspired by energy in the spin-based models which
captures the degree of alignment between connected individuals. The `local'
social energy that an individual $n\in \EuScript{N}$ feels can be defined by:
\begin{eqnarray}
E_n^{(s)}=-\frac{1}{k_{\textup{max}}\binom{M}{2}}\sum_{q \in \Gamma(n)}^{} \vec{S}_n \cdot \vec{S}_q,
\end{eqnarray}
where the sum is taken over the set of $n$'s neighbors in $\EuScript{N}$, denoted by $\Gamma(n)$.
$\vec{S}$ is a belief state vector where each element corresponds to an edge in
the belief network, so $|\vec{S}|=\binom{M}{2}$. $k_{\textup{max}}$ is a normalization constant that bounds the
strength of peer-influence and is equal to the maximum degree of $\EuScript{N}$. 
Alternatively, it could be replaced by a function that specifies the scaling 
relationship between exposure and the individual's energy.
 For our simulations everyone possesses the same set of concepts (nodes)
so $M$ is the same for each person.

We combine the internal energy with the social energy for all individuals to
define the total energy as follows: 
\begin{equation}\label{eq:ham}
%\begin{split}
%
H = \sum_{n \in \EuScript{N}} \left[ J E_n^{(i)} + I E_n^{(s)} \right] %\\
%&= -J \sum_{n \in \EuScript{N}} \left[ \frac{1}{M_{\Delta}} \sum_{j,k,l} a_{jk} a_{kl}
%a_{jl} \right] - I \sum_{n,q \in \EuScript{N}, n \neq q} \vec{S}_n \cdot \vec{S}_q, 
%
%\end{split}
\end{equation}
where the last sum is taken over all nodes on the network. 
The parameters $J$ and $I$, which we refer to as the
\emph{coherentism} and \emph{peer-influence} respectively, control the relative
contribution of the internal energy and the social energy to the total. 
The dynamics is dominated by internal belief coherence if $J \gg I$ and by social 
consensus when $I \gg J$.

Each individual is endowed with their own internal belief network and may
transfer some of their beliefs to their social contacts.  A receiver of a
belief either accepts the incoming belief or not based on the context of their
own belief system (internal coherency) and similarity to their neighbors
(social conformity).  A belief is more likely to be accepted if it increases
the coherence of an individual's own belief system, social pressure will also
increase the odds of a belief being accepted, even if it conflicts with their
belief system.

We implement these ideas by creating the following rules: at each time step
$t$, a random pair of connected individuals is chosen and one of the
individuals (sender) randomly chooses a belief (association) from its internal
belief system and sends it to the other individual (receiver), as illustrated
in Fig.~\ref{Fig 1}(c).  We assume that each individual has an identical
set of concept nodes.   Figure~\ref{Fig 1}(c) shows the selection and
emission process on a graph. The receiver accepts the association if it decreases their individual energy: 
$H_n=JE^{(i)}_n+IE^{(s)}_n$.
Even if the change in energy is less than zero, $\Delta H_n>0$, the receiver may 
still accept with the probability of $e^{\frac{-\Delta H_n}{T}}$. This term is analogous to
the Boltzmann factor~\cite{schroeder-thermal-2000}.  $T$, which we refer to as
\emph{susceptibility}, serves a similar purpose as temperature in physical
systems for the belief network.  As $T$ increases, an individual is more likely
to accept their neighbor's opinions that conflict with their own.

We characterize the status of the whole society by defining two global energy
functions. First, the \emph{mean individual energy} $\langle E^{(i)} \rangle$
measures the average internal coherence of individuals.  It is expressed by the
following equation:
\begin{eqnarray}\label{eq:avgindivE}
\langle E^{(i)} \rangle= \frac{1}{N}\sum_{n \in \EuScript{N}}
E^{(i)}_{n}.
\end{eqnarray}
The average is taken over the energies of all individuals and it can take
values between $+1$ and $-1$.  $\langle E^{(i)} \rangle = -1$ means that every
individual in the society possesses a completely coherent belief system, with
no pressured beliefs.  The other extreme ($+1$) represents a society where
every individual has completely incoherent beliefs.

Yet, this measure does not give us any indication of how homogeneous a society
is, as belief systems can vary widely while still being coherent. We have a
second energy measure inspired from spin systems:
\begin{eqnarray}\label{eq:e_s}
\langle E^{(s)}\rangle = \frac{1}{N}\sum_{n \in \EuScript{N}} E_n^{(s)},
\end{eqnarray}
which is minimized if the society is in consensus.

For each simulation we use Erd\"{o}s-R\`{e}nyi graphs with $N=10^4$ nodes and average degree of $5$, though similar results are found for 2D lattices. The belief network was fully connected with $M=5$.

\section*{Results}

Most opinion models exhibit a phase transition from a disordered state to an
ordered one~\cite{castellano-socdyn-2009}, where the ordered state represents
consensus.  As our model includes the two conflicting forces---personal belief
rigidity and social influence---we first ask how the relative strength of these
two forces governs consensus dynamics.

\begin{figure*}
\centering
	\includegraphics[width=17.8cm]{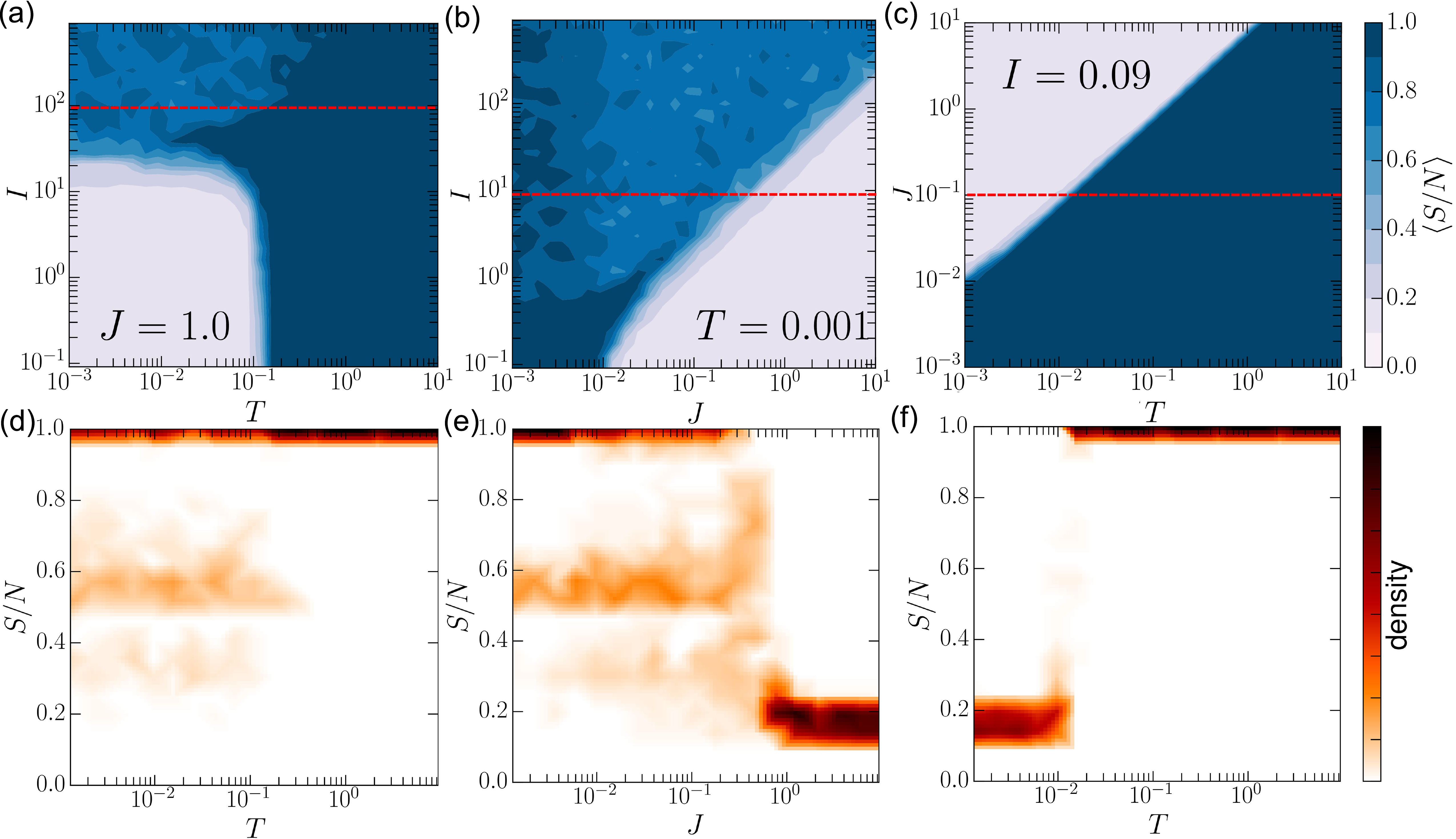}

\caption{\label{Fig 2} \textbf{Phase space} (a-c) Phase diagrams of various combinations of the
three parameters $J$, $I$, and $T$. Along with corresponding slices through
the phase space as indicated by the dashed red lines (bottom row).  (a,d) peer-influence $I$ and susceptibility $T$
conflict creating a regime where multiple belief systems with various
coherences can coexist. We see a similar regime appear in (b, e) where
peer-influence and coherentism contend for dominance. More traditional
disorder-to-order transitions as in other opinion models also take place when
$I$ is small and fixed (c, f). $S/N$ is the fractional size of the largest group.
ER graphs with $N=10^4$ nodes and average degree of $5$ were used. The density was calculated from a $160$ trails per point. 
The belief network was fully connected with $M=5$.}
\end{figure*}

Through social interaction, the society may or may not reach a consensus,
depending upon the relative strength of peer-influence ($I$) and coherentism
($J$).  Figure~\ref{Fig 2} shows a set of phase diagrams for various
combinations of $J$, $I$, and $T$. $S/N$ is defined as the belief system with 
the largest number of constituents normalized by the size of the social network. Density in Figure~\ref{Fig 2}(d)-(f) is the probability density of $S/N$ over $160$ trials at a given parameter configuration.
Since belief systems evolve locally,
individuals may have to pass through incoherent states before fully converting.
As the coherentism ($J$) increases individuals tend to cling to their own
beliefs rather than make such incoherent transitions. This can prevent
consensus, but small local consensus can still occur.  As the susceptibility
($T$) increases individuals readily accept incoherent beliefs, allowing
individuals to traverse through the belief space. It facilitates spreading of
ideas and consensus. By making individuals more susceptible to belief spreading
rather than to random switching, $T$ plays the opposite role to temperature in
standard spin models~\cite{castellano-socdyn-2009}. $T$ acts as a temperature
for individual's beliefs, and as an inverse-temperature for the whole system.
As the peer-influence ($I$) increases individuals again become more prone to
consensus. Novel dynamics occur when the effect of coherentism and social
influence become comparable.  Figure~\ref{Fig 2}(b) shows that complete
consensus is not guaranteed when the forces exerted by $J$ and $I$ remain
comparable in size. The competition leads to a situation where belief systems
can coexist and where less coherent systems can dominate.
In Figure~\ref{Fig 2}(d)-(e), we see examples of these competing configurations for when T or J is decreased and I held constant, as denoted by the higher density of largest belief-systems around 0.5. In these cases there are usually two (sometimes more) larger competing belief systems that cannot completely dominate the system even after a very long time (hundreds of billions of time-steps). As noted in~\cite{marvel-social_balance_energy-2009} there are local minima distributed throughout belief space where an individual's belief system can get stuck. At lower temperatures and when the drive for internal coherence (J) is much smaller than I, then groups of individuals (sometimes the whole population) can collectively become stuck in these “jammed” states. Under these conditions convergence to a completely coherent belief system is not guaranteed.

In our model the energy contribution of belief networks can have a major impact
on system stability as individuals seek more coherent beliefs and resolve
dissonance. What happens when key beliefs are upset through an external shock
or perturbation?

\begin{figure}[!ht]
	\includegraphics[width=8.0cm]{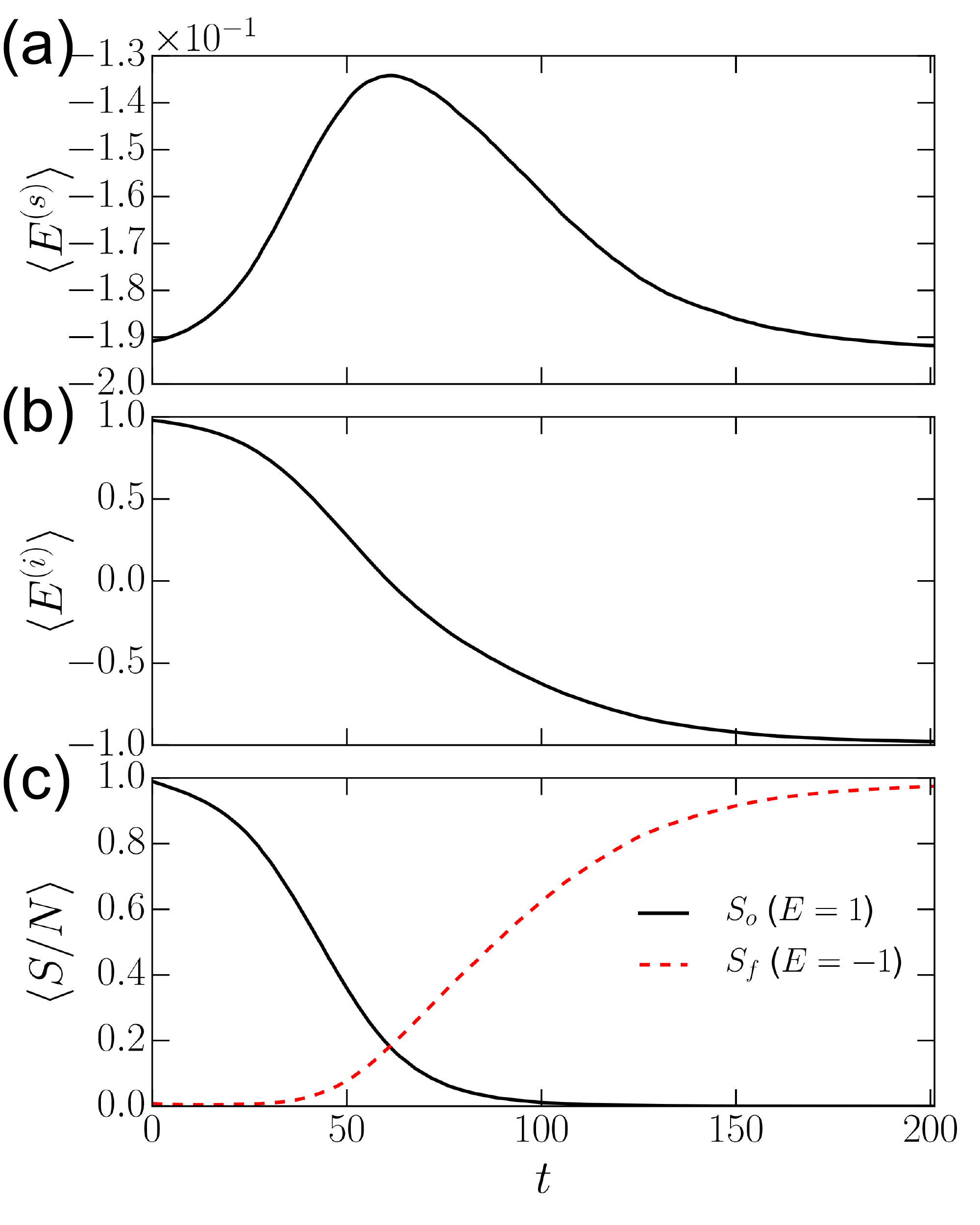}

\caption{\label{Fig 3}  \textbf{Belief driven social instability.} Strong
societal consensus does not guarantee a stable society in our model. If major
paradigm shifts occur and make individual
belief systems incoherent, then society may become unstable. (a)
The plot shows the evolution of social energy $E^{(s)}$ over time. The system
starts at consensus but with incoherent beliefs. After introducing a small perturbation, 
individuals leave consensus, searching for more coherent sets of beliefs, until society re-converges at a
stable configuration. (b) Decreasing mean individual energies $\langle
E^{(i)} \rangle$ over time illustrates individual stabilization during societal
transition.  (c) $\langle S/N \rangle$ is the fractional group size. 
As society is upset, the original dominant but incoherent belief system $S_o$
(solid black) is replaced by an emerging coherent alternative $S_f$ (dashed red).  }

\end{figure}

Imagine two independent systems, both homogeneous with the same social energy,
$E^{(s)}$. In traditional social models, with the absence of the individual
belief system (cognitive factor), these systems are identical.  By contrast,
the internal system of interconnected beliefs introduces a new force that
drives people to seek coherence in the structure of their own belief systems.
Given a homogeneous population of people with \emph{highly coherent} belief
systems, society remains stable. However, given a homogeneous population of
\emph{incoherent} belief systems, society will become unstable and following a small perturbation, 
breaks down (see Fig.~\ref{Fig 3}). In our simulation, the society is initialized at consensus with an incoherent belief system. Then 1\% of the population are given a random belief system. Individuals attempt to reduce the energy of their own belief systems and leave consensus. This society eventually re-converges at a more coherent belief-system that is different from the original consensus (Fig.~\ref{Fig 3}(c)). In the model, consensus does not guarantee stability.

These social instabilities may help explain why new ideologies can arise in the presence of belief systems that dominate most of the population. Traumatic events such as war or depression could make previously coherent beliefs less coherent, thereby destabilizing the belief system as a whole. Since these beliefs are shared by the overwhelming majority, this perturbation has the potential to have widespread impact. Such changes may
reduce the internal coherence of individuals, which may make the whole
society more prone to paradigm shifts. Throughout
history, major external perturbations in the form of war, depression, and
crippling inflation has frequently disrupted the world-views of a society's
citizens, inducing subsequent social upheavals. More coherent belief systems,
which otherwise fail to gain adherents in the presence of a dominant stable
societal values, could then gain the upper-hand by recruiting among a population of
disturbed citizens as Hoffer suggests~\cite{hoffer-truebeliever-1966}.

Given that the coherence of personal beliefs fundamentally impacts collective
behavior, we investigate the impact of ``true believers'' or zealots in a
population. They can play an important role in shaping collective social dynamics~\cite{hoffer-truebeliever-1966, 
castellano-socdyn-2009}. 
Zealots were introduced in~\cite{Mobilia-Zealot-2003} and~\cite{Mobilia-voting-2004} in the context of the voter model. Since, there has been continued work on the impact of zealots in the voter model~\cite{Redner-zealots-2007, Kashisaz-voter-2014, Modilia-qvoterzealots-2015} as well as binary adoption~\cite{Galehouse-opinions-2014}, naming game~\cite{dsouza-namingzealots-2015, Mistry-namingzealot-2015}, and other social models~\cite{Xie-minorities-2011, Arendt-zealots-2015}. The related topic of minority spreading has also been explored in various opinion models~\cite{Galam-minority-2002, Tessone-minority-2004, Alvarez-Galvez-minority-2015}.
Here we explore this aspect of opinion dynamics from the perspective of cognitive forces acting on the agents.
In the context of our framework we define zealots as individuals
who will never alter their own belief systems, but will continue to attempt to
convert others to their own.

\begin{figure}[!ht]
	\includegraphics[width=8.4cm]{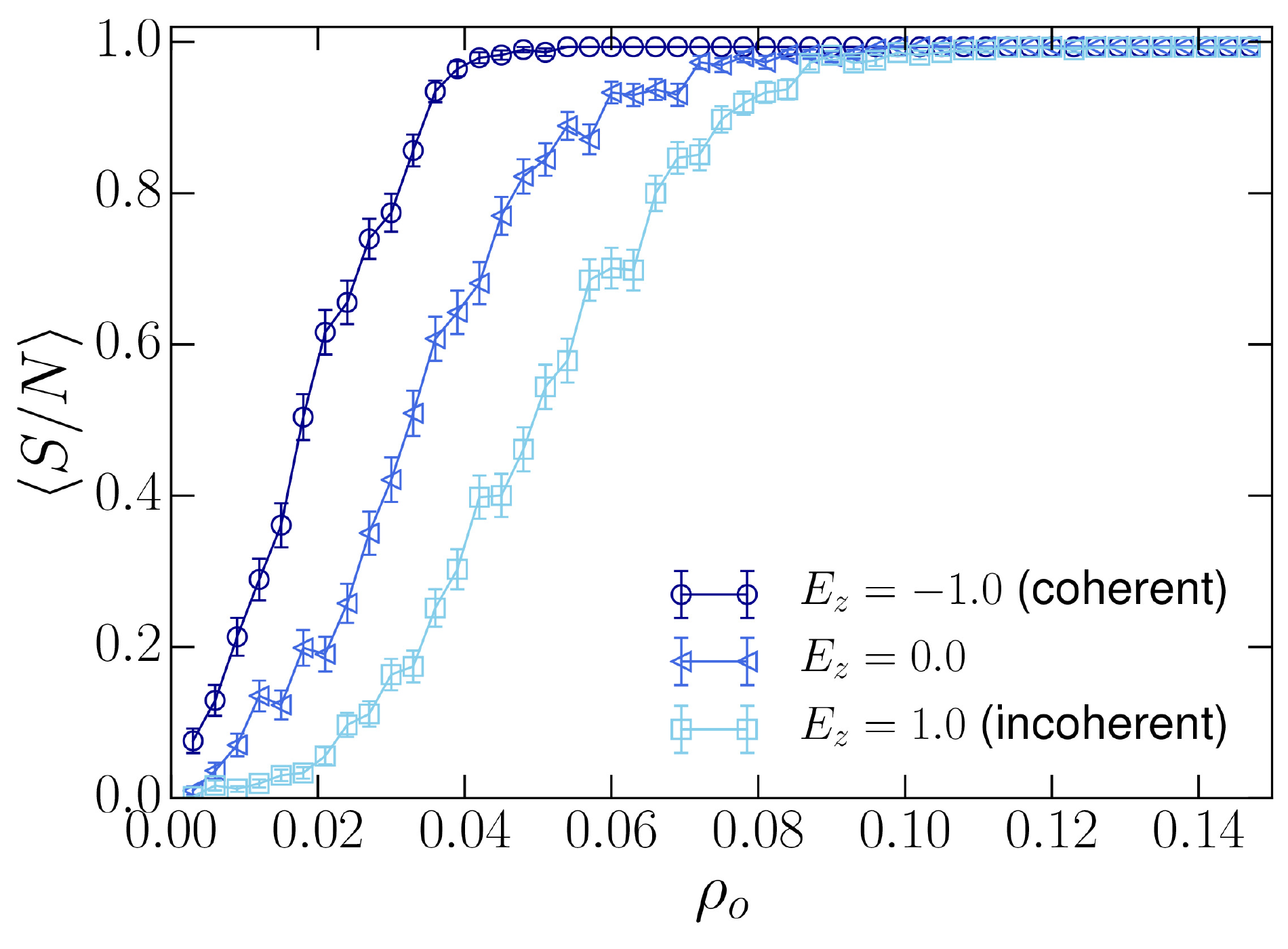}

\caption{\label{Fig 4} \textbf{Impact of the internal consistency of
zealot beliefs.} We define zealots as a group of individuals who share an
identical, immutable belief system. Such belief systems can, however, vary in
terms of their coherency.  The dynamics of $\langle S / N \rangle$, the fractional size of
the zealot population, over
$\rho_o$, the density of zealots introduced into the population, reveals that
zealots with more coherent beliefs can convert a population much more
efficiently.  In converting the whole population, the coherent set of beliefs
(circles) require only less than half the density of zealots compared with
incoherent beliefs (squares). Bars show standard error and $E_z$ is the energy of
the zealot's belief-system. The simulations were run using $J=2.0$, $T=2.0$ and $I=90.0$. }

\end{figure}

To study the impact of zealots, we prepare a homogeneous society with highly
coherent beliefs and introduce zealots with varying internal coherence.
As seen in studies of previous social models, there is a \emph{tipping-point} density
above which the minority opinion takes over the population.
Figure~\ref{Fig 4} shows that the internal coherence of zealots has a
strong impact on their effectiveness in converting society. Low coherence
zealots require much higher densities in order to convert the whole population
(see squares in Fig.~\ref{Fig 4}) because converted individuals revert
back to more coherent belief-systems at a higher rate, making it difficult for
the zealot's belief-system to retain converts.  Highly coherent zealots pull
the whole population out of consensus and convert it to their belief-systems
more easily.  Coherent zealots require almost half the density, in this
particular case, to convert the entire population. This suggests that coherence
of a set of beliefs plays a vital role in determining how well the set of
beliefs spread through society.

Our model may offer a possible explanation for the
co-existence of many seemingly invalid or impractical cults and fringe groups
in our society. The beliefs of cults and other fringe groups may frequently
contradict reality, yet they continue to thrive in spite of being surrounded by
large majorities of other belief systems.  In our model, this is explained by
the degree to which the coherence of belief systems can out-balance social
pressure in addition to social isolation.

\begin{figure*}
	\includegraphics[width=11.4cm]{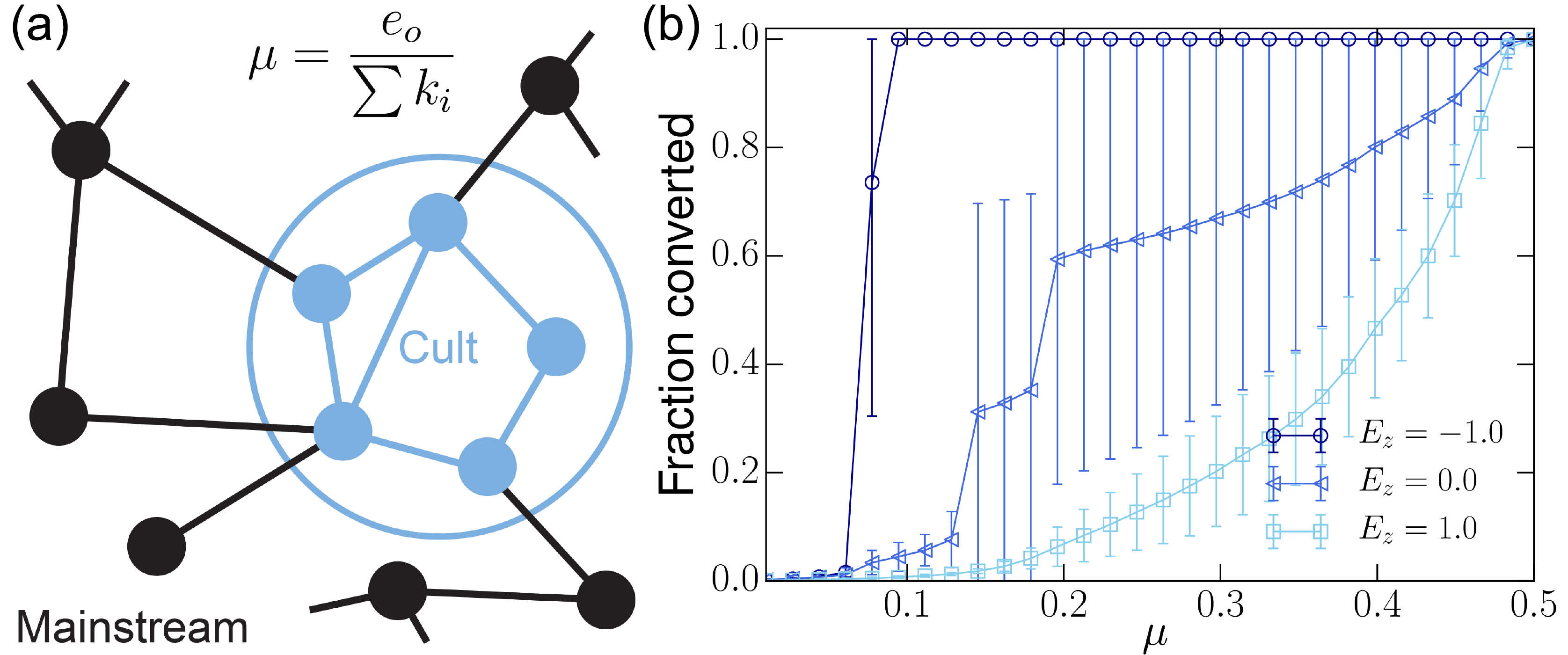}

\caption{\label{Fig 5} \textbf{Belief Invasion} (a) The survival of cults and fringe groups
depends on the coherence and strength of beliefs. We create a network with two
communities with parameters $T=2.0$, $I=0.09$, and $J=2.0$---putting the
system in a regime where it will seek consensus.  We vary the fraction of links
that connect the cult community to the mainstream community, denoted $\mu$.
$e_o$ is the number of social links between communities and $\sum k_i$ is the
total number of links in the cult (both shared and internal).  The mainstream
community attempts to convert the smaller cult.  (b) At low $\mu$ the lack of
exposure allows the cult to resist mainstream conversion.  At higher $\mu$
there is sufficient exposure to the mainstream community to overcome the
rigidity of the cult's belief system.  However, the process of conversion
becomes more difficult as the cult's beliefs become more coherent than
mainstream beliefs.  Cults are easily converted with highly coherent mainstream
beliefs even at low exposure levels (black circles), while cults maintain their
beliefs even at high exposure given low coherence of mainstream beliefs (red
squares). Bars show standard deviation. }

\end{figure*}

To investigate the influence of belief rigidity and social isolation on social
dynamics we create a society with two communities, a large mainstream community
and a smaller cult community (Fig.~\ref{Fig 5}(a)).  The
\emph{mainstream} community acts like a reservoir of mainstream beliefs.  The
other community consists of people following a different coherent belief
system.  By controlling social exposure and the strength of belief coherence,
we investigate what effect belief coherence has on the capacity for mainstream
society to invade the cult.  The parameter $\mu$ controls the fraction of edges
in the cult community that are shared with the mainstream
community~\cite{lancichinetti-lfr_graphs-2008}.  When socially isolated (low
$\mu$) the cult can resist invasion regardless of mainstream coherence
(Fig.~\ref{Fig 5}(b)). Groups of like-minded individuals can resist
outside influence by reducing social contact with non-members (decreasing
$\mu$) and enhancing internal social interactions, both of which are common in
many fringe groups and even in religious communities. 
On the other hand, the internal coherence also plays a key role. Less coherent
mainstream beliefs have greater difficulty in converting the cult, even at high
levels of exposure (high $\mu$). Compared with the dogma of cults, the truth or
the reality can be more complex and less coherent. In such case, even a
belief-system firmly grounded on truth may struggle to convert cults that
possess a highly coherent set of beliefs.  Interlocked beliefs turns
tightly-knit communities into bastions of resistance to ideological invasion
from outside.

\begin{figure}[h!]
	\includegraphics[width=8.4cm]{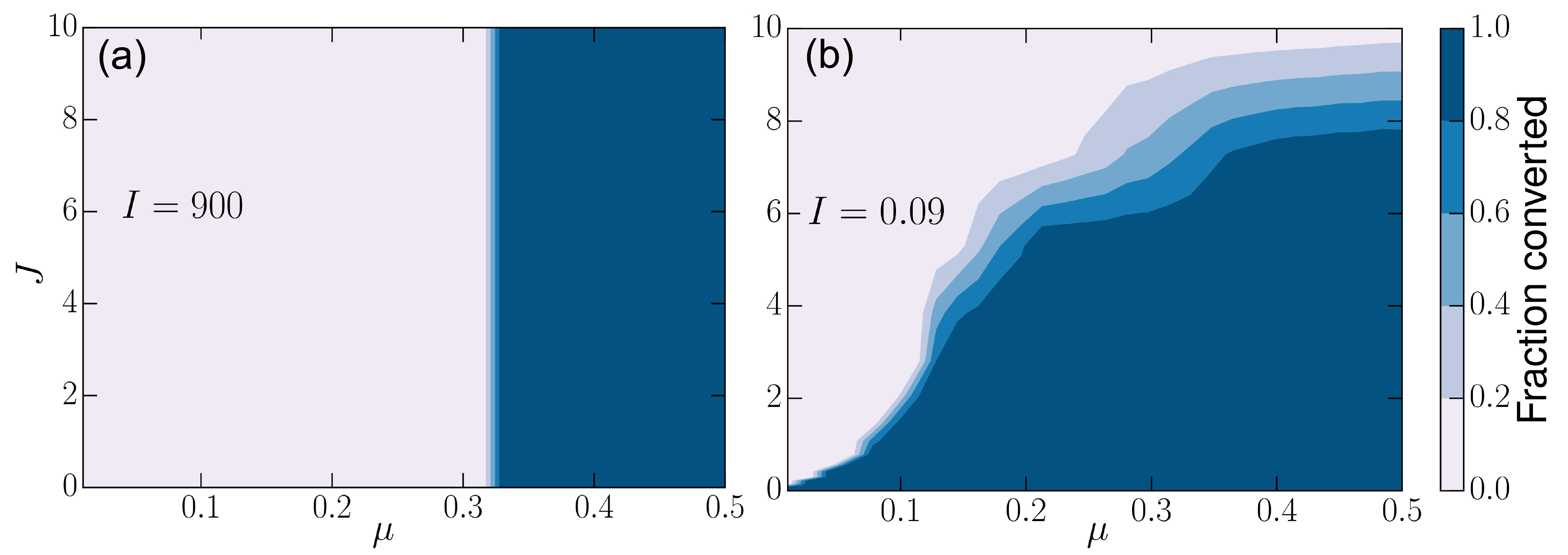}

\caption{\label{Fig 6} \textbf{Community and belief rigidity} (a) Exposure determines conversion resistance
when peer-influence ($I$) is strong.  (b) Fringe groups can sustain their
beliefs, even at a very high level of social exposure, with high levels of
individual coherentism ($J$).  }

\end{figure}

Figure~\ref{Fig 6}(a) shows a phase diagram of the case when $I \gg J$,
where conversion is determined solely by social exposure ($\mu$). This is
expected by most spin-based social models.  However, if the strength of social
influence is comparable to belief strength, social exposure is not the sole
determinant anymore (see Fig.~\ref{Fig 6}(b)). Notably, by increasing
the coherentism ($J$) of its members and maintaining a coherent set of beliefs,
cults can continue to thrive even with \emph{complete mixing} with society.
This contradicts, while underlining common intuition, the traditional exposure
models where enough exposure is sufficient to convert populations. These
results also hint at common characteristics of surviving cults. We do not
expect to find successful cults that have low belief coherence and high mixing
because they would be quickly converted by the mainstream society. We expect to
see more cults that utilize a combination of policies that minimize their
member's social contacts with outsiders, emphasize the importance of their
dogma (increasing $J$), and maximizing the coherence of their beliefs. The
latter could mean incorporating explanations for beliefs that contradict
empirical evidence or belittling mainstream methods of reasoning.  Coherence
can be, but is not necessarily aligned with logical consistency, rather
coherence is based on the strength of associations between concepts and valence
concepts (see SKS model~\cite{greenwald-iat-2002}), that is derived from a
connectionist cognitive framework.

\section*{Discussion}

We have shown that our model exhibits a disorder-to-order transition similar to
other opinion models, while also exhibiting unique dynamics that could help explain common
processes observed in the real world, such as the breakdown of a
homogeneous society driven by shocks to individuals' beliefs, the dependence of
zealotry on belief coherence, and the successful entrenchment of
fringe groups.  Each of these results arises from the fact that our framework
integrates belief interdependence with social influence.

The breakdown of homogeneous societies can manifest through changes in the
connectivity of concepts at the concept-network level (cf. SKS model). Abrupt
shocks to these conceptual connections can impact the coherence of beliefs.
That impact is reflected in our model as an unstable belief system. When people
share a common shock, one could expect that many of those people's
belief-systems will enter a frustrated state. In order for this socially
frustrated state to be resolved, members of the population partake in
societal-dissonance resolution strategies by collectively transitioning toward
more coherent belief systems. From this view we could interpret paradigm shifts in culture as something akin to a societal-wide coping strategy.

Our model is a minimalistic approach toward introducing individualist
concept-networks into a systematic social dynamics model.  Though simple, the
framework can be expanded to include more realistic features.  For instance,
communicated beliefs are chosen randomly, but this feature can be replaced with
other models of belief communication, such as imitation, where someone will be
more likely to communicate recently accepted beliefs.  Additionally, our
mapping from the SKS concept model could be expanded to included weighted
relationships that emphasize the uncertainty that an individual has in their
beliefs (conviction) and its relations to other beliefs.  While we used the
same parameter values ($J$, $I$, and $T$) for all individuals, we expect them to vary within the
population from person to person.  Variations of these parameters could produce the
natural occurrence of zealot-like behavior. Finally, the formation and maintenance of 
social ties is an important facet of social dynamics~\cite{lazer-indivnet-2001, uzzi_socialties_2010}
and could be implemented by allowing agents to choose their neighbors~\cite{Cook-homophily-2001}.

Our model expands upon previous social models by implementing an internal
belief system that assumes that the interdependence among beliefs shapes their
dynamics. In this framework we are able to incorporate known psychological
forces into the model and show that a bottom-up approach can successfully link
micro-level behaviors to global dynamics. By making this small jump to using
internal belief-systems, while preserving key features of standard social
models, such as percolation and global consensus, it extends their dynamics and explanatory value, and could be a contributing factor for explaining the entrenchment of belief systems and social upheaval. Future work will be directed towards the development of new
opinion models for specific applications that more closely integrate our
behavioral and cognitive understanding of humanity.

\section*{Acknowledgments}
The authors would like to thank Alessandro Flamini, Rob Goldstone, and Fabio
Rojas for valuable feedback and discussions.

%\section*{References}
% Either type in your references using
% \begin{thebibliography}{}
% \bibitem{}
% Text
% \end{thebibliography}
%
% OR
%
% Compile your BiBTeX database using our plos2015.bst
% style file and paste the contents of your .bbl file
% here.
% 

%\bibliography{references}

\begin{thebibliography}{10}

\bibitem{sowell-ideology_conflict-2002}
Sowell T.
\newblock A conflict of visions: ideological origins of political struggles.
\newblock New York: Basic Books; 2002.

\bibitem{hoffer-truebeliever-1966}
Hoffer E.
\newblock The true believer: thoughts on the nature of mass movements.
\newblock Perennial library. New York: Harper and Row, 1966, c1951.; 1966.

\bibitem{rohkramer-germany-2007}
Rohkramer T.
\newblock A single communal faith?: the German Right from Conservatism to
  National Socialism.
\newblock New York: Berghahn Books; 2007.

\bibitem{mcdonough-hitler-2012}
McDonough F.
\newblock Hitler and the Rise of the Nazi Party.
\newblock Harlow, England: Pearson; 2012.

\bibitem{britannica-reformation-2014}
Britannica E. History of Europe; 2014.

\bibitem{axelrod-culture-2009}
Axelrod R.
\newblock The Dissemination of Culture: A Model with Local Convergence and
  Global Polarization.
\newblock The Journal of Conflict Resolution. 2009;42(2):203--226.

\bibitem{hammond-ethnocentrism-2006}
Hammond RA, Axelrod R.
\newblock The Evolution of Ethnocentrism.
\newblock The Journal of Conflict Resolution. 2006;50(6):926.

\bibitem{castellano-socdyn-2009}
Castellano C, Fortunato S, Loreto V.
\newblock Statistical physics of social dynamics.
\newblock Reviews Of Modern Physics. 2009;81(2):591--646.

\bibitem{Masuda-voter-2010}
Sood V, Antal T, Redner S.
\newblock Voter models on heterogeneous networks.
\newblock Phys Rev E. 2008 Apr;77:041121.
\newblock Available from:
  \url{http://link.aps.org/doi/10.1103/PhysRevE.77.041121}.

\bibitem{sood-hetnetvoter-2007}
Sood V, Antal T, Redner S.
\newblock Voter models on heterogeneous networks.
\newblock Phys Rev E. 2008 Apr;77:041121.
\newblock Available from:
  \url{http://link.aps.org/doi/10.1103/PhysRevE.77.041121}.

\bibitem{sznajd-opinionmodel-2000}
Sznajd-Weron K, Sznajd J.
\newblock Opinion Evolution in Closed Community.
\newblock International Journal of Modern Physics C. 2000;11(6):1157.

\bibitem{clifford-votermodel-1973}
Clifford P, Sudbury A.
\newblock A Model for Spatial Conflict.
\newblock Biometrika. 1973;60(3):581.

\bibitem{Javarone-conformity-2014}
Javarone MA.
\newblock Social influences in opinion dynamics: the role of conformity.
\newblock Physica A: Statistical Mechanics and its Applications.
  2014;414:19--30.

\bibitem{Galam-opinion-2000}
Galam S, Zucker JD.
\newblock From individual choice to group decision-making.
\newblock Physica A: Statistical Mechanics and its Applications.
  2000;287(3–4):644--659.

\bibitem{Galam-review-2008}
Galam S.
\newblock Sociophysics: A Review of Galam Models.
\newblock International Journal of Modern Physics C. 2008;19(03):409--440.

\bibitem{Guillaume-opinion-2000}
Guillaume D, David N, Frederic A, Gérard W.
\newblock Mixing beliefs among interacting agents.
\newblock Advances in Complex Systems. 2000;3(1-4):87.

\bibitem{Hegselmann-opinion-2002}
Hegselmann R, Krause U.
\newblock Opinion dynamics and bounded confidence: models, analysis and
  simulation.
\newblock JASSS-The Jouranl of Artificial Societies and Social Simulation.
  2002;5(3).

\bibitem{Allahverdyan-ConfirmationBias-2014}
Allahverdyan A, Galstyan A.
\newblock Opinion Dynamics with Confirmation Bias.
\newblock PLoS One. 2014;9(7):1 -- 14.

\bibitem{nickerson-confirmationbias-1998}
Nickerson R.
\newblock Confirmation bias: A ubiquitous phenomenon in many guises.
\newblock Review of General Psychology. 1998;2(2):175 -- 220.

\bibitem{festinger-dissonance-1962}
Festinger L.
\newblock A theory of cognitive dissonance; 1962.

\bibitem{goldstone-collectiveBehavior-2005}
Goldstone R, Janssen M.
\newblock Computational models of collective behavior.
\newblock Trends in cognitive sciences. 2005;9(9):424 -- 430.

\bibitem{mason-socialpsycreview-2007}
Mason WA, Conrey FR, Smith ER.
\newblock Situating social influence processes : Dynamic, multidirectional
  flows of influence within social networks.
\newblock Personality and social psychology review. 2007;11(3):279 -- 300.

\bibitem{mcgarty-turnerexp-1992}
McGarty C, Turner JC.
\newblock The effects of categorization on social judgement.
\newblock British Journal of Social Psychology. 1992;31:253–268.

\bibitem{pechey-sharedbeliefs-2012}
Pechey R, Halligan P.
\newblock Exploring the Folk Understanding of Belief: Identifying Key
  Dimensions Endorsed in the General Population.
\newblock Journal of Cognition and Culture. 2012;12(1/2):81 -- 99.

\bibitem{redner-truth-2011}
Masuda N, Redner S.
\newblock Can partisan voting lead to truth?
\newblock Journal of Statistical Mechanics: Theory and Experiment.
  2011;2011(02):L02002.
\newblock Available from:
  \url{http://stacks.iop.org/1742-5468/2011/i=02/a=L02002}.

\bibitem{nishi-socialconfbias-2013}
Nishi R, Masuda N.
\newblock Collective opinion formation model under Bayesian updating and
  confirmation bias.
\newblock Physical Review E. 2013;87(6).

\bibitem{Wang-opinion-2014}
Wang S, Huang C, Sun C.
\newblock Modeling self-perception agents in an opinion dynamics propagation
  society.
\newblock Simulation-Transactions of The Society For Modeling and Simulation
  International. 2014;90(3):238 -- 248.

\bibitem{javarone-socialstructure-2013}
Javarone M, Armano G.
\newblock Perception of similarity: a model for social network dynamics.
\newblock Journal of Physics A. 2013;46(45).

\bibitem{Sekara-socialstructure-2015}
Sekara V, Stopczynski A, Lehmann S.
\newblock Fundamental structures of dynamic social networks.
\newblock Proceedings of the National Academy of Sciences. 2016;Available from:
  \url{http://www.pnas.org/content/early/2016/08/22/1602803113.abstract}.

\bibitem{shi-coevolvingvotermodel-2013}
Shi F, Mucha PJ, Durrett R.
\newblock Multiopinion coevolving voter model with infinitely many phase
  transitions.
\newblock Physical Review E - Statistical, Nonlinear, and Soft Matter Physics.
  2013;88(6).

\bibitem{lord-attitude_polarize-1979}
Lord C, Ross L, Lepper M.
\newblock Biased assimilation and attitude polarization: The effects of prior
  theories on subsequently considered evidence.
\newblock Journal of Personality and Social Psychology. 1979;37(11):2098 --
  2109.

\bibitem{taber-politicalbel-2006}
Taber C, Lodge M.
\newblock Motivated Skepticism in the Evaluation of Political Beliefs.
\newblock American Journal of Political Science. 2006;50(3):755 -- 769.

\bibitem{evans-beliefsys-xxxx}
Evans E.
\newblock Cognitive and contextual factors in the emergence of diverse belief
  systems: Creation versus evolution.
\newblock Cognitive Psychology. 2001;42(3):217 -- 266.

\bibitem{greenwald-iat-2002}
Greenwald A, Banaji M, Rudman L, Farnham S, Nosek B, Mellott D.
\newblock A unified theory of implicit attitudes, stereotypes, self-esteem, and
  self-concept.
\newblock Psychological Review. 2002;109(1):3 -- 25.

\bibitem{schilling-cognet-2005}
Schilling MA.
\newblock A ``small-world" network model of cognitive insight.
\newblock Creativity Research Journal. 2005;17(2-3):131--154.

\bibitem{quine-beliefweb-1970}
Quine W, Ullian J.
\newblock The web of belief.
\newblock New York, Random House; 1970.

\bibitem{thagard-coherence-2000}
Thagard P.
\newblock Coherence in thought and action.
\newblock Life and mind. Cambridge, Mass. : MIT Press; 2000.

\bibitem{pechey-beliefcoherence-2012}
Pechey R, Halligan P.
\newblock Using Co-Occurrence to Evaluate Belief Coherence in a Large Non
  Clinical Sample.
\newblock PLoS One. 2012;7(11):1 -- 8.

\bibitem{heider-cog_organization-1946}
Heider F.
\newblock Attitudes and Cognitive Organization.
\newblock Journal of Psychology. 1946;21:107--112.

\bibitem{norman-balancetheory_group-2003}
Norman H, Patrick D.
\newblock Some dynamics of social balance processes: bringing Heider back into
  balance theory.
\newblock Social Networks. 2003;25:17 -- 49.

\bibitem{asch-groupress-1951}
Asch SE.
\newblock Effects of group pressure on the modification and distortion of
  judgments. In Groups, leadership and men; 1951.

\bibitem{asch-conformity-1956}
Asch SE.
\newblock Studies of independence and conformity. A minority of one against a
  unanimous majority.
\newblock Psychological Monographs. 1956;70(9):1–70.

\bibitem{centola-onlinenet-2010}
Centola D.
\newblock The Spread of Behavior in an Online Social Network Experiment.
\newblock Science. 2010;5996:1194.

\bibitem{matz-cogdis_in_groups-2005}
Matz D, Wood W.
\newblock Cognitive Dissonance in Groups: The Consequences of Disagreement.
\newblock Journal of Personality and Social Psychology. 2005;88(1):22 -- 37.

\bibitem{heider-original-1946}
Heider F.
\newblock Attitudes and Cognitive Organization.
\newblock The Journal of Psychology. 1946;21(1):107--112.

\bibitem{antal-social_balance-2006}
Antal T, Krapivsky P, Redner S.
\newblock Social Balance on Networks: The Dynamics of Friendship and Enmity.
\newblock Physica D. 2006;224:130--136.

\bibitem{doreian-social_balance-2002}
Doreian P.
\newblock Event sequences as generators of social network evolution.
\newblock Social Networks. 2002;24(2):93.

\bibitem{cartwright-balance_theory-1957}
Cartwright D, Harary F.
\newblock Structural balance: a generalization of Heider's theory.
\newblock Psychological Review. 1956;63(5):277 -- 293.

\bibitem{elliot-cogdis-1994}
Elliot A, Devine P.
\newblock On the motivational nature of cognitive dissonance: Dissonance as
  psychological discomfort.
\newblock Journal of Personality and Social Psychology. 1994;67(3):382 -- 394.

\bibitem{marvel-social_balance_energy-2009}
Marvel S, Strogatz S, Kleinberg J.
\newblock Energy Landscape of Social Balance.
\newblock Physcial Review Letters. 2009;103(19).

\bibitem{schroeder-thermal-2000}
Daniel S.
\newblock Boltzmann Statistics.
\newblock In: Iwata S, editor. Introduction to Thermal Physics. Addison Wesley
  Longman; 2000. .

\bibitem{Mobilia-Zealot-2003}
Mobilia M.
\newblock Does a Single Zealot Affect an Infinite Group of Voters?
\newblock Phys Rev Lett. 2003 Jul;91:028701.
\newblock Available from:
  \url{http://link.aps.org/doi/10.1103/PhysRevLett.91.028701}.

\bibitem{Mobilia-voting-2004}
Mobilia M, Georgiev IT.
\newblock Voting and catalytic processes with inhomogeneities.
\newblock Phys Rev E. 2005 Apr;71:046102.
\newblock Available from:
  \url{http://link.aps.org/doi/10.1103/PhysRevE.71.046102}.

\bibitem{Redner-zealots-2007}
Mobilia M, Petersen A, Redner S.
\newblock On the role of zealotry in the voter model.
\newblock Journal of Statistical Mechanics - Theory and Experiment. 2007;p.~17.

\bibitem{Kashisaz-voter-2014}
Kashisaz H, Hosseini SS, Darooneh AH.
\newblock The effect of zealots on the rate of consensus achievement in complex
  networks.
\newblock Physica A: Statistical Mechanics and its Applications. 2014;402:49 --
  57.
\newblock Available from:
  \url{http://www.sciencedirect.com/science/article/pii/S0378437114000764}.

\bibitem{Modilia-qvoterzealots-2015}
Mobilia M.
\newblock Nonlinear $q$-voter model with inflexible zealots.
\newblock Phys Rev E. 2015 Jul;92:012803.
\newblock Available from:
  \url{http://link.aps.org/doi/10.1103/PhysRevE.92.012803}.

\bibitem{Galehouse-opinions-2014}
Galehouse D, Nguyen T, Sreenivasan S, Lizardo O, Korniss G, Szymanski B.
\newblock Impact of network connectivity and agent commitment on spread of
  opinions in social networks.
\newblock Proceedings of the 5th International Conference on Applied Human
  Factors and Ergonomics. 2014;p. 2318--2329.

\bibitem{dsouza-namingzealots-2015}
Waagen A, Verma G, Chan K, Swami A, D'Souza R.
\newblock Effect of zealotry in high-dimensional opinion dynamics models.
\newblock Phys Rev E. 2015 Feb;91:022811.
\newblock Available from:
  \url{http://link.aps.org/doi/10.1103/PhysRevE.91.022811}.

\bibitem{Mistry-namingzealot-2015}
Mistry D, Zhang Q, Perra N, Baronchelli A.
\newblock Committed activists and the reshaping of status-quo social consensus.
\newblock Phys Rev E. 2015 Oct;92:042805.
\newblock Available from:
  \url{http://link.aps.org/doi/10.1103/PhysRevE.92.042805}.

\bibitem{Xie-minorities-2011}
Xie J, Sreenivasan S, Korniss G, Szymanski BK, Zhang W, Lim C.
\newblock Social consensus through the influence of committed minorities.
\newblock Physical Review E - Statistical, Nonlinear, and Soft Matter Physics.
  2011;84(1).

\bibitem{Arendt-zealots-2015}
Arendt DL, Blaha LM.
\newblock Opinions, influence, and zealotry: a computational study on
  stubbornness.
\newblock Computational and Mathematical Organization Theory. 2015;21(2):184.

\bibitem{Galam-minority-2002}
Galam S.
\newblock Minority opinion spreading in random geometry.
\newblock European Physical Journal B. 2002;25(4):403 -- 406.

\bibitem{Tessone-minority-2004}
Tessone CJ, Toral R, Amengual P, Wio HS, Miguel MS.
\newblock Neighborhood models of minority opinion spreading.
\newblock European Physical Journal B. 2004;39:535--544.

\bibitem{Alvarez-Galvez-minority-2015}
Alvarez-Galvez J.
\newblock Network Models of Minority Opinion Spreading: Using Agent-Based
  Modeling to Study Possible Scenarios of Social Contagion.
\newblock Social Science Computer Review. 2015;34(5):567--581.

\bibitem{lancichinetti-lfr_graphs-2008}
Lancichinetti A, Fortunato S.
\newblock Benchmarks for testing community detection algorithms on directed and
  weighted graphs with overlapping communities.
\newblock Physical Review E. 2008;80(1).

\bibitem{lazer-indivnet-2001}
Lazer D.
\newblock The co-evolution of individual and network.
\newblock Journal of Mathematical Sociology. 2001;25(1):69 -- 108.

\bibitem{uzzi_socialties_2010}
Rivera MT, Soderstrom SB, Uzzi B.
\newblock Dynamics of dyads in social networks: Assortative, relational, and
  proximity mechanisms.. vol.~36 of Annual Review of Sociology; 2010.

\bibitem{Cook-homophily-2001}
McPherson M, Smith-Lovin L, Cook J.
\newblock Birds of a Feather: Homophily in Social Networks.
\newblock Annual Review of Sociology. 2001;p. 415.

\end{thebibliography}

\end{document}